# Resonant Thermoelectric Nanophotonics

Kelly W. Mauser[1], Slobodan Mitrovic[2], Seyoon Kim[1], Dagny Fleischman[1], and Harry A. Atwater[1,3,*]

* haa@caltech.edu

1. Thomas J. Watson Laboratory of Applied Physics, California Institute of Technology, Pasadena, CA 91125, United States

2. Joint Center for Artificial Photosynthesis, California Institute of Technology, Pasadena, CA 91125, United States

2. Kavli Nanoscience Institute, California Institute of Technology, Pasadena, CA 91125, United States

**Photodetectors are typically based on photocurrent generation from electron-hole pairs in semiconductor structures and on bolometry for wavelengths that are below bandgap absorption. In both cases, resonant plasmonic and nanophotonic structures have been successfully used to enhance performance. In this work, we demonstrate subwavelength thermoelectric nanostructures designed for resonant spectrally selective absorption, which creates large enough localized temperature gradients to generate easily measureable thermoelectric voltages. We show that such structures are tunable and are capable of highly wavelength specific detection, with an input power responsivity of up to 119 V/W (referenced to incident illumination), and response times of nearly 3 kHz, by combining resonant absorption and thermoelectric junctions within a single structure, yielding a bandgap-independent photodetection mechanism. We report results for both resonant nanophotonic bismuth telluride – antimony telluride structures and chromel – alumel structures as examples of a broad class of nanophotonic thermoelectric structures useful for fast, low-cost and robust optoelectronic applications such as non-bandgap-limited hyperspectral and broad-band photodetectors.**

Plasmon excitation enables extreme light confinement at the nanoscale, localizing energy in subwavelength volumes and thus can enable increased absorption in photovoltaic or photoconductive detectors[1]. Nonetheless, plasmon decay also results in energy transfer to the lattice as heat which is detrimental to photovoltaic detector performance[2]. However, heat generation in resonant subwavelength nanostructures also represents a power source for energy conversion, as we demonstrate here via design of resonant thermoelectric (TE) plasmonic absorbers for optical detection. Though TEs have been used to observe resonantly coupled surface plasmon polaritons in noble-metal thin films and microelectrodes[3,4] and have been explored theoretically for generation of ultrafast intense magnetic pulses in a dual-metal split ring resonator[5], they have not been employed as resonant absorbers in functional TE nanophotonic structures. Previously, non-narrowband photodetection has been demonstrated through the photothermoelectric effect in gated graphene structures[6,7] and the laser heating of nanoantennas and micropatterned materials[8-22], all shown to be promising for infrared to teraherz broadband detection. Typical responsivities of the graphene structures are around 10 V/W for IR and THz detectors, relative to incident (not absorbed) power, with a time response ranging from 23 ms to nearly 10 ps. Responsivities of non-graphene detectors range from 10's of V/W to nearly 2,400 V/W[13] for thermopiles made of many thermocouples. The response time of these structures range from 10's to 100's of ms, though GHz response times have been predicted[8] for nanoantenna structures. High-figure-of-merit TE's have been investigated as solar power generators, but the light absorption process was entirely separate from the TE functionality and relied on black carbon absorbers[23] or solar concentrators[24].

We propose and demonstrate here nanostructures composed of TE thermocouple junctions using established TE materials – chromel/alumel and bismuth telluride/antimony telluride – but patterned so as to support guided mode resonances (GMRs) with sharp absorption profiles, and which thus generate large thermal gradients upon optical excitation and localized heat generation in the TE material. Unlike the TE absorbers described above, they feature tunable narrowband absorption and measured single junction responsivities 10 times higher than the most similar graphene structures[6,22], with potential for much higher responsivities in thermopile architectures. For bismuth telluride – antimony telluride structures, we measure thermoelectric voltages (TEVs) up to 850 µV with incident optical power densities of 3.4 W/cm$^2$. The maximum responsivity of a single thermocouple structure was measured at 119 V/W, referenced to incident illumination power. We also find that the small heat capacity of optically resonant TE nanowires enables a fast, 337 µs temporal response, 10-100 times faster than conventional TE detectors. We note that TE nanostructures have also been shown to display improved performance[25-29] by increased phonon boundary scattering in high surface-to-volume ratio geometries with dimensions comparable to phonon mean-free-paths; while we do not believe that our structures at present exploit this phenomenon, this represents an opportunity for future nanophotonic TE structures. We show that TE nanophotonic structures are tunable from the visible to the MIR, with small structure sizes of 50 µm x 100 µm. Our nanophotonic TE structures suspended on a thin membranes to reduce substrate heat losses and improve thermal isolation between TE structures arranged in arrays suitable for imaging or spectroscopy. Whereas photoconductive and photovoltaic detectors are typically insensitive to sub-bandgap radiation, nanophotonic TEs can be designed to be sensitive to any specific wavelength dictated by nanoscale geometry, without bandgap wavelength cutoff limitations. From the point of view of imaging and spectroscopy, they enable integration of filter and photodetector functions into a single structure.

**POWER FLOWS IN THERMOELECTRIC PLASMONIC STRUCTURES**

Figure 1a depicts power flows in a nanophotonic TE structure, and Figs. 1b,c shows a schematic of our experimental structure, a guided mode resonance (GMR) wire array, in which optical radiation is coupled into a waveguide mode via a periodic TE wire array that serves as a light absorber. Optical power is generated at the TE junction, while the ends of the TE wires are at ambient temperature, resulting in a thermoelectric voltage (TEV). Our nanophotonic TE structures on membrane substrates have dimensions large enough that bulk heat transport equations can be used (i.e. no ballistic or quantized thermal conductance). The steady state temperature of the illuminated region is found by balancing light absorption with energy loss via radiation, conduction through the interface, conduction to the unilluminated material, and convection to the surrounding gas ambient:

$$P_{absorbed} = P_{radiated} + P_{interface} + P_{conducted} + P_{convected}$$

To characterize the responsivity, we seek to determine the TEV, proportional to the Seebeck coefficient, $S$, and the temperature difference $\Delta T$, between cold and hot ends of the material, *i.e.,* $TEV = S\Delta T$.

At steady state, the balance of absorbed power and dissipated power is then

$$\int_{V_{TE}} \frac{1}{2}\omega\varepsilon''|\boldsymbol{E}|^2 dV$$

$$= \int_{A_{air}} e\sigma(T^4 - T_0^4)dA - \int_{A_{slice}} \kappa\nabla T dA + \int_{A_{air}} h(T - T_0)dA$$

$$+ \int_{A_{interface}} h_c(T - T_{sub})dA, \quad (1)$$

where $\omega$ is the incident light frequency, $\varepsilon''$ is the imaginary part of the TE material dielectric function, $|\boldsymbol{E}|$ is the magnitude of the electric field within the TE material, $V_{TE}$ is the volume of the illuminated TE material, $e$ is the TE material emissivity, $A_{air}$ is the area of the TE exposed to the air, $\sigma$ is the Stefan-Boltzmann constant, $T_0$ is the initial temperature (also cold end temperature/ambient air temperature), $\kappa$ is the TE material thermal conductivity, $A_{slice}$ is the TE material cross-sectional area separating the illuminated and unilluminated regions, $h$ is the heat transfer coefficient between the TE material and air, $h_c$ is the thermal boundary conductance between the TE material and the substrate, $T_{sub}$ is the temperature of the substrate near the TE material, and $A_{interface}$ is the area of the intersection of the TE material with the substrate. A structure optimized for maximum responsivity (minimize power lost from the illuminated region and maximize power absorbed into the illuminated region) would be a perfectly absorbing structure that is non-emissive in directions other than the light absorption direction, suspended in vacuum. Our design is a compromise among these factors, with a temperature increase over ambient $\Delta T$ of approximately 1K calculated via this simple model (see Supplementary note 1).

As an example of a TE plasmonic nanostructure, we consider a periodic array of wires composed of TE materials on a thin, suspended, electrically insulating, low thermal conductivity substrate. A wire array/substrate heterostructure supporting GMRs in an n/p-type TE junction is shown in Fig. 1c. An alumel/chromel junction structure is shown in Fig. 1d, with wires 100 nm in width, 40 nm in height, and 470 nm period, fabricated on a 145 nm thick freestanding dielectric slab waveguide composed of 45 nm SiO$_2$ and 100 nm SiN$_x$ layers, and this structures yields the resonant absorption spectrum depicted in Fig. 1e. The absorption resonance can be shifted by several hundred nm by varying the wire array period. Thus a periodic tiling of wire array pixels each with a different period and resonance frequency could function as a TE hyperspectral detector, shown conceptually in Fig. 1f.

**PHOTONIC DESIGN OF RESONANT THERMOELECTRIC STRUCTURES: LINE SHAPE AND PEAK POSITION**

Nanophotonic TE structures must concentrate the electric field in the TE material to maximize absorption. Our GMR structures acheive this via Fano interference[30] of a waveguide mode and a Fabry-Perot resonance in the waveguide (c.f., Supplementary Figs. 1-3, Supplementary equations (S2-S7)). The location of this waveguide mode is predicted quite well by the grating coupler equation for normally incident light, assuming infinitely narrow gratings, $\frac{2\pi}{d} = \beta$, where $d$ is the grating pitch and $\beta$ is the propagation constant of the two-layer slab waveguide; while interaction with the Fabry-Perot resonance shifts the waveguide resonance, this a small correction.

A wide range of materials with varying Seebeck coefficients including Al, Cr, antimony telluride, and doped Si give rise to GMRs with very similar peak heights, positions, and widths, as shown in Fig. 2a (see Supplementary Fig. 6 for the dielectric function of antimony telluride). Antimony telluride and Cr exhibit large extinction $k$ at the waveguide resonance wavelength and are plasmonic ($\varepsilon'<0$) in this wavelength range; Al, which has a larger magnitude negative $\varepsilon'$ in this region has a narrower resonant linewidth, whereas Au, Cu, and Ag have resonances shifted due to interband transitions or plasmon resonances that couple to the waveguide mode causing a Rabi splitting of the modes[31].

Cross-sections of bismuth telluride wire GMR structures are shown in Figs. 2b-g, with 40 nm height, 68 nm width and 488 nm period on top of a 50 nm SiO2/100 nm SiN$_x$ waveguide layer, all suspended in air. Figures 2b,d,f,h correspond to the absorption maxima wavelength, and Figs. 2c,e,g correspond to the absorption minima just to the left of the maxima, as shown in Fig. 2i. Figure 2b shows the electric field surrounding the wires at the maximum absorption wavelength, resulting from a constructive interference of the waveguide mode and the Fabry-Perot resonance. The large electric field magnitude in the wire corresponds to high power absorption on resonance, shown in Fig. 2d (inset is enlarged wire), whereas Fig. 2c illustrates the off-resonance electric field, at an absorption minimum, shown in Fig. 2e. Figures 2f and g are two-dimensional simulations that indicate the wire array temperature distributions on and off resonance, also indicating a large temperature difference between these states. Figure 2h is a three-dimensional thermal simulation of an array of antimony telluride/bismuth telluride wire junctions under Gaussian illumination with a 1 µm beam waist. This simulation indicates a locally elevated temperature at the junction and small temperature difference between center and edge of our structure, which sustains a temperature gradient in steady state. For computational reasons, the 3 µm x 5 µm simulation field is smaller than the 50 µm by 100 µm field in our experimental structures, which are expected to sustain a larger temperature gradient between hot and cold regions. Figure 2i shows the absorption and corresponding wire temperature spectral distribution obtained from the two-dimensional simulations. We see that the temperature spectral distribution does not exactly follow the absorption spectrum, which we interpret as nonlinear wire heat loss with temperature, since the radiative power term (see equation (1)) depends on the fourth power of temperature.

TE nanophotonic structures supporting GMRs exhibit tunable narrowband absorption over a wide wavelength range by variation of wire array geometrical parameters. We can tune the absorption resonance over the entire visible spectrum at constant waveguide thickness (50 nm $SiO_2$, 100 nm $SiN_x$) by varying the wire array period, as shown in in Fig. 3a, for the structure depicted in Fig. 3c. The peak spectral position is thus mainly dictated by the wire array period as well as waveguide thickness and material indices, and only slightly by wire material. Variation of wire width results in a small peak shift, but primarily increases the peak absorption and linewidth FWHM, as seen in Figs. 3a,b. We find that increased wire height increases peak absorption up to a height of 40 nm, whereas absorption asymptotes for greater thicknesses (see Supplementary Fig. 7). Figure 3d shows the experimental extinction (black line), simulated absorption (red line), and simulated best-fit (blue dotted line) for parameters listed in Supplementary Table 2. Absorption (red line, dimensions extracted from sample SEM images) was calculated using full wave simulations and $P_{abs} = \frac{1}{2}\omega\varepsilon''|\mathbf{E}|^2$, normalized to the incident power input in the simulation. The measured absorption (black line) was determined from absorption = (1 – transmittance – reflectance) (described in Methods). The peak positions in our experiment closely match those predicted by simulations, including the minor peak to the short wavelength side of the main absorption peak, which is due to the off-normal incidence illumination and the angle sensitivity of GMR structures (see Supplementary Fig. 7). The incident illumination angle used in simulation, either 0.5 or 1 degree off normal incidence, was chosen based on the best fit to the data. The general predictability of these peak locations using simulations makes fabrication of a hyperspectral pixel much simpler. The most noticeable discrepancy between our simulation and experiment is the FWHM of the peaks. Our experiment performed better than simulation in creating narrow FWHM peaks for use in a hyperspectral detector. This points to potentially a geometrical cause, such as not perfectly rectangular wires or surface roughness. The best-fit simulation (blue dotted line) was achieved by fitting the experimental data with altered wire dimensions in simulations. Values for the simulation and best-fit simulation can be found in Supplementary Table 2. The best-fit simulation wire width was thinner than the experimentally measured wire width by between 0-29 nm. Fitting experimental and simulation spectra to a Fano shape[32] for one wire pitch (described in Supplementary Figs. 1-5, Supplementary Table 2), we found that the experimental spectra exhibited larger damping caused by losses in our structure, which altered the absorption spectra shape.

The absorption maxima can be tuned across several hundred nanometers of wavelength for a given waveguide thickness. To access IR waveguide modes which produce large absorption peaks, the waveguide thickness must be increased. Figure 3e shows wavelength versus wire pitch for a 50/100 nm $SiO_x/SiN_x$ waveguide thickness. This configuration is ideal for sensing visible to NIR wavelengths. Again, we can determine the approximate absorption peak location in our system using $\frac{2\pi}{d} = \beta$. As shown in Fig. 3f, by selecting a 300/500 nm thick $SiO_2/SiN_x$ layer, we can shift the absorption peaks beyond the detection limit of Si photodetectors, which is around 1100 nm. We can create absorption peaks in the MIR around 4-5 μm if we adjust the waveguide spacing to 500/500 nm $SiO_2/SiN_x$ and increase the pitch to several microns, as shown in Fig. 3g. In principle, the only limitation in IR tunability for these detectors is the phonon absorption band in SiO$_2$ (and SiN$_x$) at around 8-11 µm[33,34]. Using different materials as a waveguide layer could further extend the range of these detectors.

**SPECTRAL RESPONSE AND RESPONSIVITY**

Figure 4 summarizes the measurements for our TE plasmonic GMR structures. We found the voltage to be linearly dependent on incident power, as shown in Fig. 4g for a bismuth telluride – antimony telluride structure with absorption spectra depicted by the red curve in Fig. 4c. We found weighted root mean squared error values of 0.58 µV, 0.45 µV, 1.05 µV, 0.82 µV, and 0.74 µV for our first order polynomial fit for 700 nm, 675 nm, 650 nm, 625 nm, and 600 nm, respectively, which strongly suggests linear dependence of TEV on incident power. For each wavelength, the voltage amplitude is proportional to power absorbed, i.e. a wavelength of 650 nm yields higher absorption and voltage than 600 nm wavelength for the same incident power. The temperature scale in Fig. 4g is based on an estimated Seebeck coefficient of 145 µV/K for antimony telluride and – 50 µV/K for bismuth telluride based on fabrication methods (see Supplementary note 6 for details). A maximum temperature gradient $\Delta T$ of ~ 4.5 K, depicted in Fig. 4g, was estimated based on these Seebeck coefficients and this lies in the general validity range for the analytical power balance in equation (1) (see Supplementary note 1 for explicit calculations).

Figures 4a-c show the spectral voltage response of a chromel-alumel structure (Fig. 4a) and a bismuth telluride-antimony telluride structure (Figs. 4b,c). Figures 4a,b show TE response spatially averaged over a ~400 µm$^2$ area of the wire array; these measurements have larger signal error bars, but indicate a measureable signal even if the wires are not illuminated only at the TE junction. Figure 4c shows a time-averaged response for temporally-chopped illumination over 400 periods for the same structure with a tightly focused source illuminating the TE junction. In Figs. 4a-c, spectral features of the voltage signal are seen to parallel the spectral absorption. For example, in Fig. 4c, we can see a voltage signal is produced with a FWHM of around 20 nm. Small discrepancies between absorption and voltage spectra are likely due to the sensitivity of GMR structures to illumination incidence angle (see Supplementary Fig. 7).

Alumel and chromel have larger thermal conductivities than bismuth telluride or antimony telluride, giving a lower *ΔT* upon illumination (see equation (1)), and have smaller Seebeck coefficients, resulting in a peak potential of 35 µV under 7.92 µW illumination for alumel-chromel, compared with a peak of 800 µV for bismuth telluride-antimony telluride under the same illumination. The structures shown in Fig. 4b-c are predicted to exhibit a ~ 4 K peak temperature gradient for an 800 µV TEV, assuming the above-mentioned Seebeck coefficients for bismuth telluride and antimony telluride.

In Figure 4d we indicate the responsivity of bismuth telluride – antimony telluride structures as a function of wavelength, the noise spectral density (Fig. 4e), and noise equivalent power (Fig. 4f). Compared to a similar graphene photothermoelectric structure[6] (with reference to incident light), our structure has a larger maximum responsivity, over 119 V/W, with a similar noise equivalent power range of 2.5-4.8 nW/Hz$^{1/2}$. The noise spectral density, with a range of 300-320 nV/Hz$^{1/2}$, could be further decreased by reducing the wire array resistance and better controlling the environment (e.g., eliminating noise due to temperature fluctuations from convection). Assuming a spot size equal to the detector area, this gives a maximum selective detectivity (*D*\*) of 6.05 x 10$^5$ cm-Hz$^{1/2}$/W, or detectivity (*D*) of 3.92 x 10$^8$ Hz$^{1/2}$/W. The responsivity of these structures could be increased further in a number of ways, including through thermopiling, optimizing the TE materials (e.g. increasing *S* in high-ZT tellurides by substrate heating during deposition or doping to increase the Seebeck coefficient[35,36]), or through altering any of the parameters described in equation (1), such as measuring in vacuum to eliminate convective loss and/or suspending the wires to eliminate conductive losses to the substrate.

Measurements of the response time under chopped illumination yielded time constants of 152.77 µs ± 3.05 µs and 153.38 µs ± 3.25 µs during heat up and cool down, respectively (see Figs. 4h,i). This corresponds to a 10%-90% rise time of ~337 µs, or almost 3 kHz, which is a fast enough response for many detection and imaging applications. This response time could be further reduced by use of smaller TE structures that decrease the overall heat capacity while maintaining high absorptivity.

We note that resonant nanophotonic TE structure design could be extended to a variety of TE configurations and resonant nanophotonic structures, several of which are illustrated in Fig.

5, including the GMR structures in Fig. 5a and a thermopile configuration designed to increase the TEV responsivity in Fig. 5b. For example, a 25 wire bismuth telluride-antimony telluride thermopile structure with a responsivity of 100 V/W per thermocouple under 8 µW of illumination would create a 20 mV signal, with no change in response time and a decrease in noise equivalent power. Addition of a backreflector below the dielectric layers could be employed to alter the waveguide mode profile, resulting in a perfect absorber geometry (Fig. 5c). Plasmonic bow-tie antennas[37] could also be designed to concentrate illumination at a TE junction (Fig. 5d), and resonant nanoparticle plasmonic structures could be used to induce wavelength-specific resonant absorption in an unpatterned thin planar TE sheet, as depicted in Fig. 5e. Finally, TE materials could be incorporated in an even broader array of resonant absorber designs[38-41], e.g., a split ring resonator perfect absorber[42] resonantly exciting a thin TE junction, illustrated in Fig. 5f.

## Acknowledgements
This work was supported primarily by US Department of Energy (DOE) Office of Science grant DE-FG02-07ER46405. S.K. acknowledges support by a Samsung Scholarship. The authors thank K. Schwab for discussions.

## Author contributions
K.W.M. and H.A.A. conceived the ideas. K.W.M., S.K performed the simulations. K.W.M. fabricated the samples. K.W.M., S.M. and D.F. performed measurements and K.W.M., S.M. performed data analysis. K.W.M., H.A.A. and S.M. co-wrote the paper, and H.A.A. supervised the project.

## Competing financial interests
The authors declare no competing financial interests.

## Methods
STRUCTURE FABRICATION

The TE hyperspectral detectors were fabricated as follows. On top of the waveguide layer of 100 nm thick $SiN_x$ membrane (Norcada NX7150C), the 50 nm $SiO_2$ spacer layer was deposited via PECVD (Oxford Instruments System 100 PECVD) at 350°C with 1000 mTorr pressure, with flows of 710 sccm $NO_2$ and 170 sccm 5% $SiH_4$ in $N_2$.

The structures were written via electron beam lithography (Raith EBPG 5000+ Electron Beam Writer) in a series of aligned writes, followed by deposition and liftoff. For all of the writes, a bilayer electron beam resist consisting of 50 nm 950 PMMA A2 on 200 nm 495 PMMA A4 was spun on top of the $SiO_x$ spacer layer. Each layer was baked for 5 minutes at 180°C in ambient conditions after spinning.

The order of structure fabrication was first alignment markers, then the bismuth telluride (or chromel) structure half, then the antimony telluride (or alumel) structure half, followed by contacts for wire bonding. The alignment markers and contacts were written with a dosage of 2100-2300 $\mu C/cm^2$, and a beam current of 10-50 nA. The resist was developed in a room temperature mixture of 1:3 methyl isobutyl ketone:IPA for 50 seconds, submerged in IPA, and dried with $N_2$. Using electron beam evaporation (Lesker Labline E-Beam Evaporator), 3-5 nm of Ti was deposited, then 60-70 nm of Au. Liftoff was performed by agitation in 70°C acetone for 30 minutes, followed by an IPA rinse.

The bismuth telluride and antimony telluride structures were written with the same dosage and a beam current of 500 pA. They were developed in 4°C 1:3 methyl isobutyle ketone:IPA for 60 seconds, submerged in IPA, and dried with $N_2$. $Bi_2Te_3$ and $Sb_2Te_3$ stoichiometric sputter targets (Stanford Materials) were used in a house-modified sputterer (Lesker). Base pressures were in the $10^{-6}$ Torr range, process pressure was 30 mTorr, gas flow was 100 sccm Ar, and the power was 40 W RF to sputter 40 nm of each material. Liftoff was performed in the manner described above. Alumel (Ni/Mg/Al/Si 95/2/2/1 wt%, VEM) and chromel (Ni/Cr 90/10 wt%, VEM) sputter targets were used with base pressure 3 mTorr, 20 sccm Ar, and 500 W DC.

EXTINCTION MEASUREMENTS

Extinction measurements were done at room temperature using a house-built setup including a 2W supercontinuum laser (Fianium), a monochromator, and a silicon photodiode (Newport). Transmission measurements were taken using a lock-in (Stanford Research systems Model SR830 DSP Lock-In Amplifier) with a chopper and current amplifier (HMS Elektronik current amplifier model 564). Extinction was calculated as 1 – transmittance - reflectance.

POTENTIAL MEASUREMENTS

Voltage measurements were done at room temperature with a voltage amplifier (Signal Recovery Model 5113 pre-amp) and a volt meter (Keithley 6430 sub-femtoamp remote source meter). Voltage signals were normalized to incident power, and interpolated using the verified relation that voltage scales linearly with incident power. Space averaged data was taken by scanning x versus y versus voltage over the structure for each wavelength and averaging a fixed set of pixels in the excited region, subtracted from a background voltage measurement off the structure. This allowed for adjustments in case of sample or beam drift. Time averaged data involved locating the beam at a place on the structure which produced maximum voltage for that wavelength, and using a chopper and oscilloscope with a voltage amplifier to measure hundreds of data points for each wavelength.

SIMULATIONS

Simulations were done using Lumerical FDTD Solutions[43], COMSOL Multiphysics[44] with RF and Heat Transfer Modules, and rigorous coupled wave analysis[45].

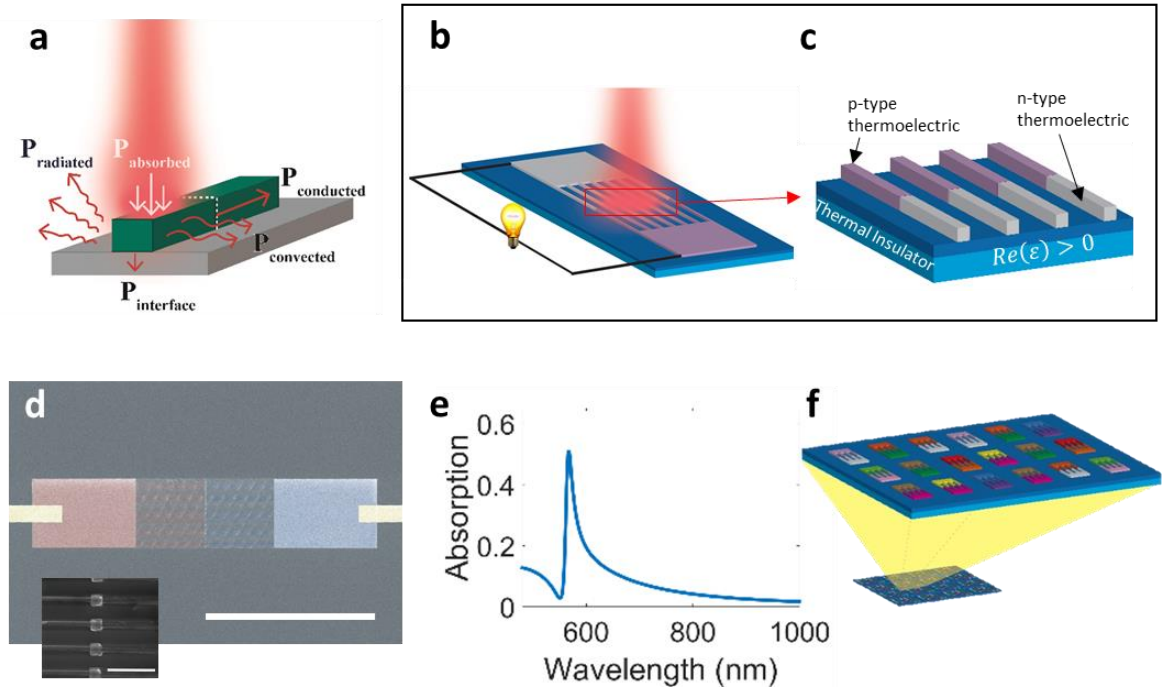

**Figure 1 | Nanophotonic power flow and guided mode resonance design. a**, Illustration of the power flow in a TE plasmonic absorbing nanstructure. **b**, Conceptual design of thermoelectric plasmonic wavelength detector. Light illuminates the junction of the thermoelectric wires and is absorbed, heating the junction and producing a thermoelectric voltage. **c**, Enlarged portion of the wire junctions in **b**. Wires sit on a suspended waveguide composed of 50 nm $SiO_2$/100 nm $SiN_x$. This plasmonic grating/waveguide structure with 40 nm high and 60 nm wide wires with a 488 nm pitch create the absorption profile shown in **e**. **d**, False color SEM of a fabricated alumel-chromel detector, with gold contacts (50 µm scale bar). Inset is the junction between bismuth telluride-antimony telluride wires in the same geometry (500 nm scale bar). Fabrication is described in Methods. **f**, Conceptual design for a hyperspectral detector pixel. By tuning the wire pitch, the absorption peak can be shifted across a several hundred nm range. By having an array of detectors of different pitches, incident light can be decomposed by wavelength.

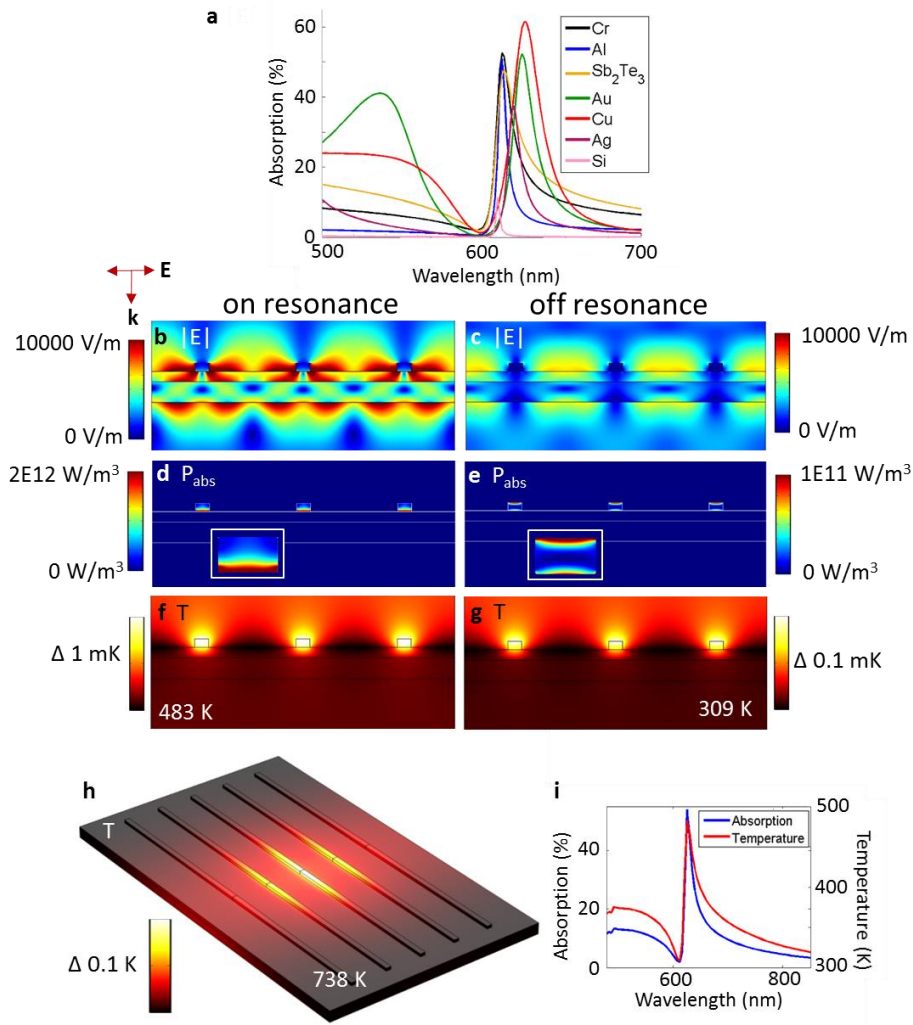

**Figure 2 | TE material performance in guided mode resonance structure design. a**, A comparison of absorption spectra of different wire materials in our GMR structure composed of 40 nm high, 68 nm wide, pitch of 488 nm wires on a waveguide of 50 nm $SiO_2$/100 nm $SiN_x$. **b-i**, 2D and 3D full wave simulations for GMR structure with dimensions of **a** with bismuth telluride wires (**h** has a bismuth telluride-antimony telluride wire junction). The structure is suspended in air. Incident light is polarized with the E-field perpendicular to the wires and at normal incidence. **b**, **d**, **f**, and **h** are at peak absorption, and **c**, **e**, and **g** are at the minimum absorption. **b-c** Magnitude of the electric field. **d-e** Absorption calculated by $P_{abs} = \frac{1}{2}\omega\varepsilon''|E|^2$, (inset is an enlarged image of the wire). **f-h** Temperature profiles. **b-g** have 1 W/cm² incident power density, while **h** has 6.8x10⁶ W/cm² incident power density. The structure shown in **g** has lateral dimensions of 3 μm by 5 μm, approximately 0.3% of the size of our fabricated structures, which explains the smaller than expected temperature gradient from experimental measurements of TEV. **I,** 2D absorption spectra and corresponding temperature for the structure simulated in **b-g**.

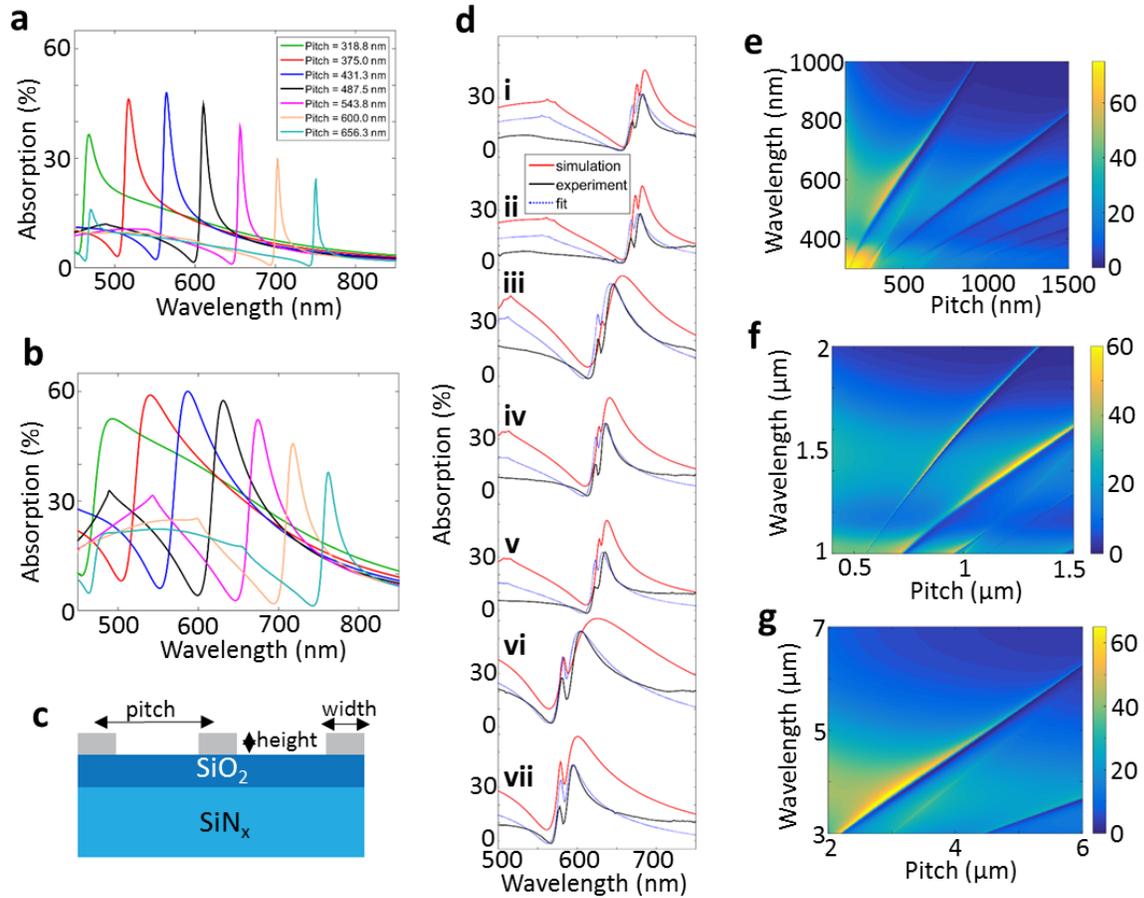

**Figure 3 | Hyperspectral absorption tunability of guided mode resonance structures: theory and experiment**. All simulations use antimony telluride or bismuth telluride for the wire material, with experimental dielectric function data given in Supplementary Fig. 6. **a**, Simulations of 60 nm wide, 40 nm high wires of varying pitch on suspended 50 nm $SiO_2$/100 nm $SiN_x$ waveguide. **b**, Configuration of **a**, but with 100 nm wide wires. **c**, 2D GMR structure geometry. **d**, Experimental extinction (black), simulated absorption corresponding to the experimental dimensions (red), and simulated absorption corresponding to fitted and scaled absorption spectra (blue dots) for varying wire pitches and widths on 45 nm $SiO_2$/100 nm $SiN_x$ waveguide (see Supplementary Table 2 for parameters). Off-normal angle of illumination causes the smaller peak to the left of the larger absorption peak to form. **e**, Wavelength versus pitch absorption plot in the visible regime for 40 nm high antimony telluride wires, on 50 nm $SiO_2$/100 nm $SiN_x$ suspended membrane. **f**, Absorption spectra for 50 nm high, 300 nm wide antimony telluride wires on 300 nm $SiO_2$/500 nm $SiN_x$. **g**, Absorption spectra in the MIR for 50 nm high, 1.5 μm wide bismuth telluride wires on 500 nm $SiO_2$/500 nm $SiN_x$.

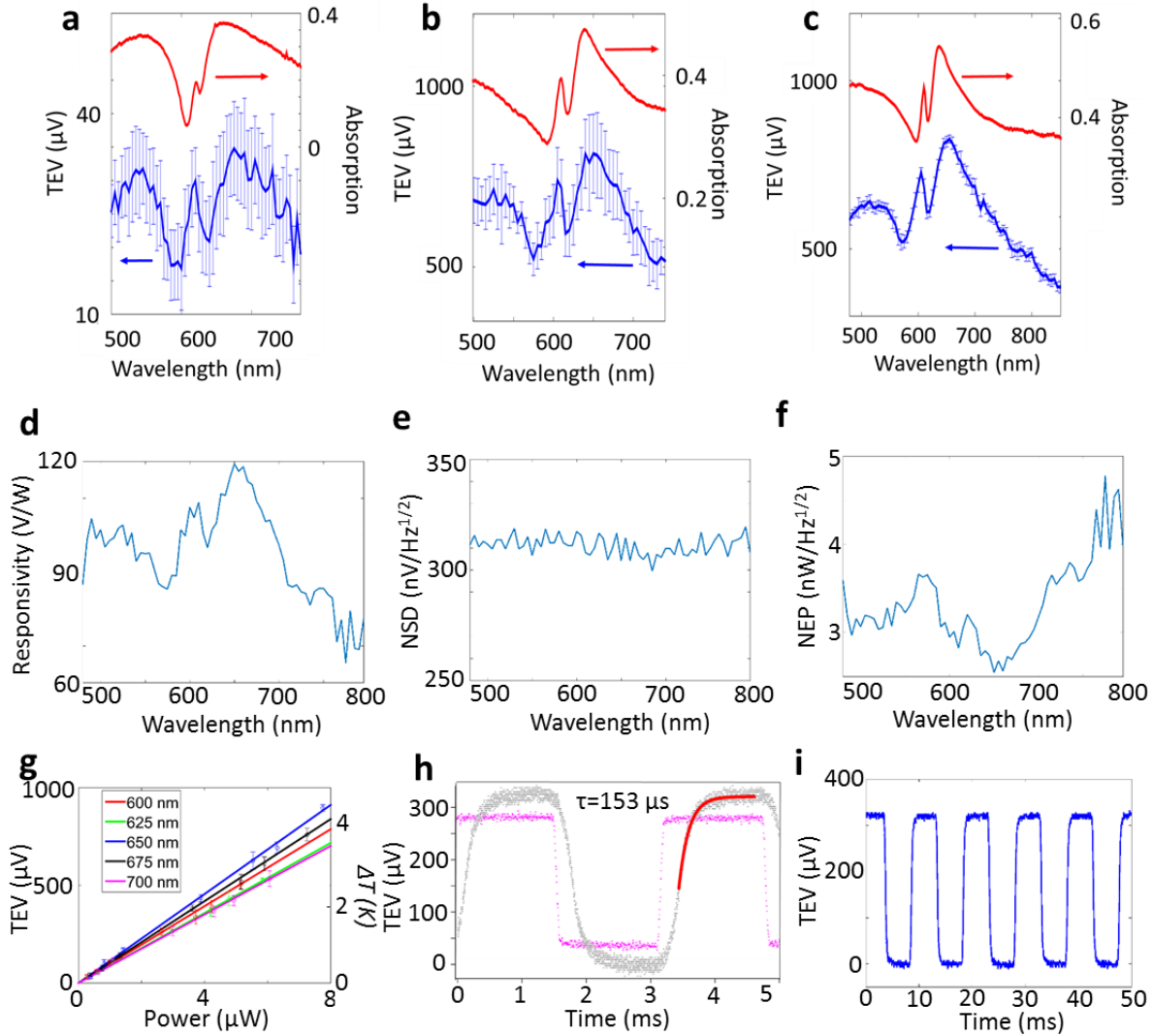

**Figure 4 | Spectral and time dependent structure performance. a**, Extinction (red) and TEV (blue) for a chromel - alumel structure under 7.92 µW illumination, or 3.4 W/cm$^2$ incident power density is shown. Each TEV data point is averaged over a ~400 µm$^2$ area. **b**, Extinction (red) and TEV (blue) for a bismuth telluride - antimony telluride structure under the same conditions as **a**. **c**, Time averaged spectral response of the bismuth telluride - antimony telluride structure shown in **b**, under the same illumination conditions. **d**-**f**, Responsivity, noise spectral density, and noise equivalent power for a bismuth telluride-antimony telluride structure. **g**, TEV dependence on incident power for a bismuth telluride - antimony telluride structure corresponding to the extinction spectra in **b**. The temperature scale on the right axis corresponds to temperature of the hot junction in wires based on the estimated Seebeck coefficient of ~195 µV/K. **h-i**, Time response of a bismuth telluride - antimony telluride structure. **h** has the time constant fit line plotted over the data from our TE detector (grey), which is measured as 152.77 µs +/- 3.05 µs, corresponding to a 10%-90% rise time of 337 µs. The pink data is the response of an Si photodiode at the same chopper speed (amplitude is unitless).

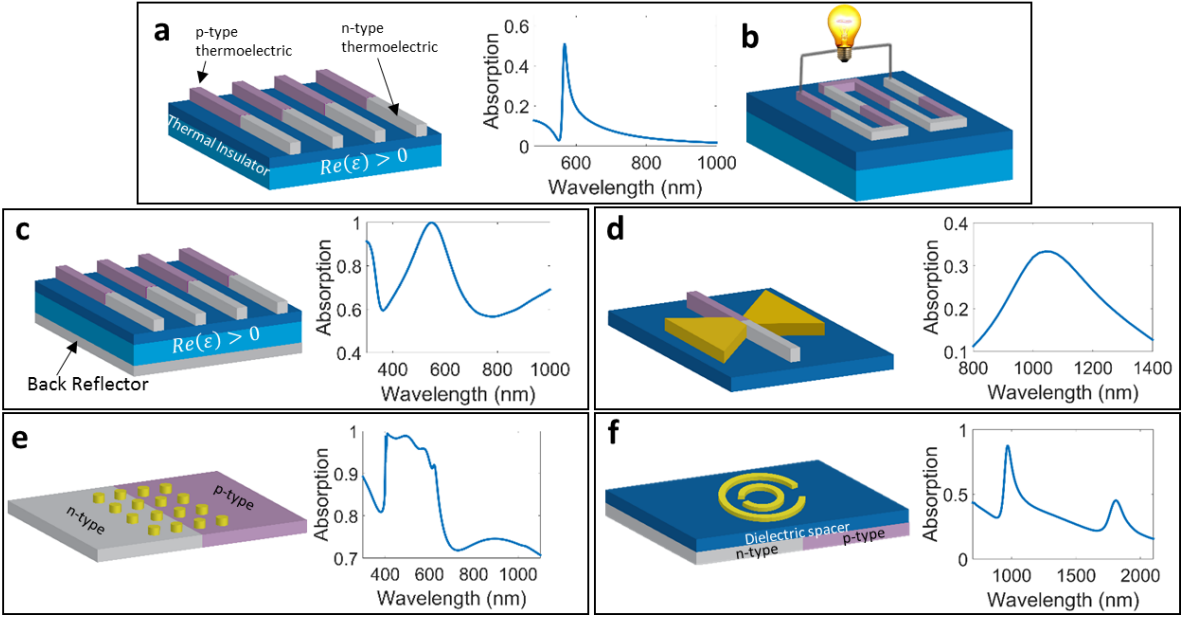

**Figure 5 | Resonantly excited nanophotonic thermoelectric structures**. **a**, GMR structure. **b**, Thermopile structure. **c**, Perfect absorber structure. **d**, Resonant bowtie structure. **e**, Resonant Mie absorbers. **f**, Split ring resonator perfect absorber.

# Resonant Thermoelectric Nanophotonics

Kelly W. Mauser[1], Slobodan Mitrovic[2], Seyoon Kim[1], Dagny Fleischman[1], and Harry A. Atwater[1,3,*]

* haa@caltech.edu

1. Thomas J. Watson Laboratory of Applied Physics, California Institute of Technology, Pasadena, CA 91125, United States

2. Joint Center for Artificial Photosynthesis, California Institute of Technology, Pasadena, CA 91125, United States

2. Kavli Nanoscience Institute, California Institute of Technology, Pasadena, CA 91125, United States

**Supplementary Notes**

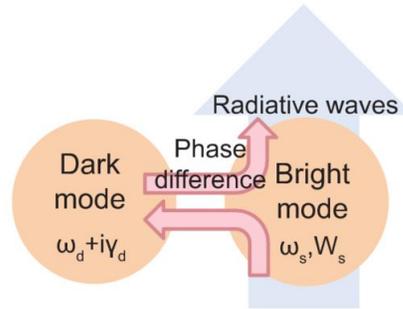

**Supplementary Figure 1 | Dark and bright modes.** Diagram of interaction of bright (broad resonance) and dark (narrow resonance) modes in the production of Fano lineshapes. From Gallinet et al.[1].

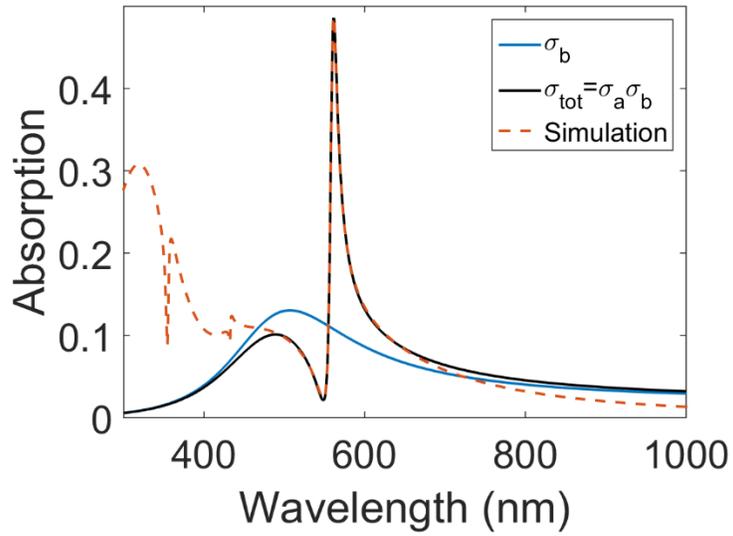

**Supplementary Figure 2 | Fano fit to simulation.** Full wave simulation of 60 nm wide, 40 nm high, pitch of 431 nm antimony telluride wires on 50 nm of $SiO_2$/100 nm of $SiN_x$ suspended waveguide fit to the Fano formula (Supplementary equation (S7)). The orange dotted line is the simulation and the black line is the fit of the combined Fano formula (Supplementary equation (S7)), between 440 nm and 650 nm. The blue line is the extracted Fano formula of the bright mode (Supplementary equation (S6)).

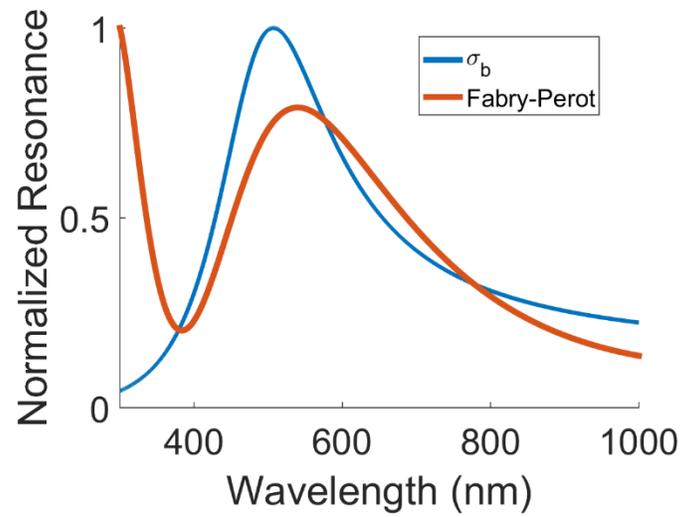

**Supplementary Figure 3 | Normalized bright mode and FP resonances.** The FP resonance is measured as the normalized magnitude of the electric field at a point on the surface of a bare waveguide structure (calculated with full wave simulations). The blue line is the bright mode fit described in Supplementary Fig. 2.

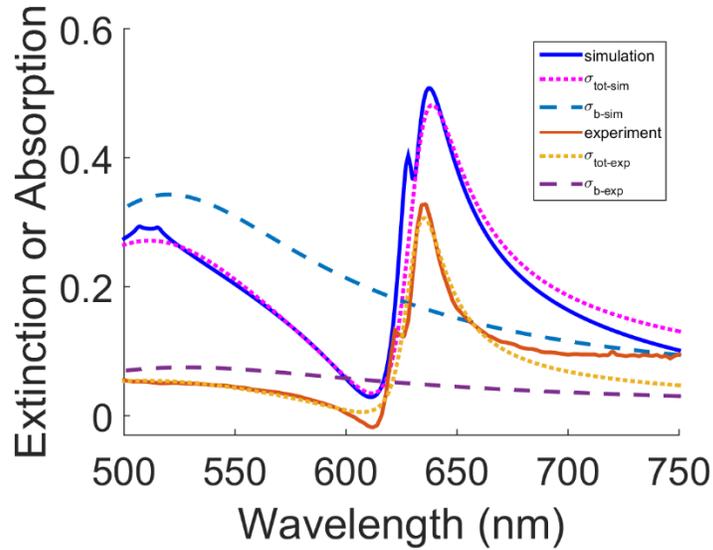

**Supplementary Figure 4 | Quantitative comparison of simulation and experiment via Fano fitting.** Experiment vs. simulation for 45 nm $SiO_2$/100 nm $SiN_x$ suspended waveguide with antimony telluride wires. Wire height is 40 nm, width is 89 nm, and pitch is 511 nm. The solid blue line corresponds to the full wave simulation for this structure, while the magenta dotted line is the Fano function fit to this line with parameters given in Supplementary Table 1. The dashed dark blue line is the bright mode extracted from the total Fano function, corresponding to the FP resonance for this structure. The solid orange line is the measured extinction. The dotted yellow line is the fitted total Fano function, and the dashed purple line is the extracted bright mode profile from the measured extinction.

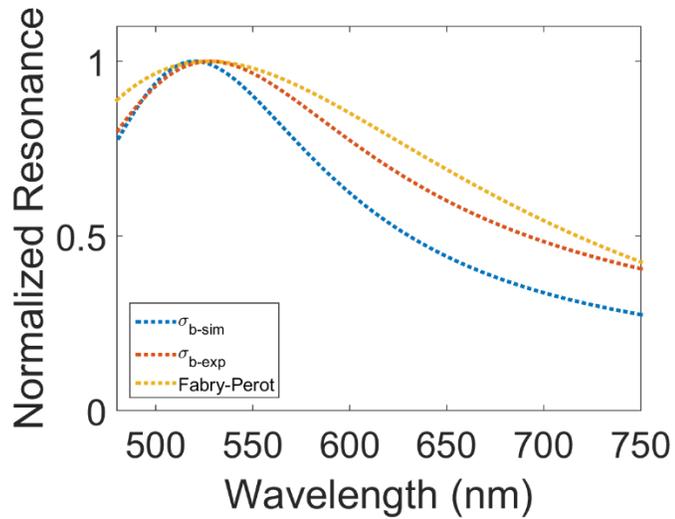

**Supplementary Figure 5 | Bright mode and FP resonance comparison**. Extracted bright mode for the experimental and simulated data from Supplementary Fig. 4, with the FP -caused electric field magnitude at the surface of the waveguide. All curves are normalized to their maximum value in the given wavelength range for ease of comparing peaks. We can see that the experimental and simulated peaks align well, and the F-P peak aligns closely as well (the FP peak will be shifted because of the contributions of the wires to the effective index of the entire photonic crystal structure).

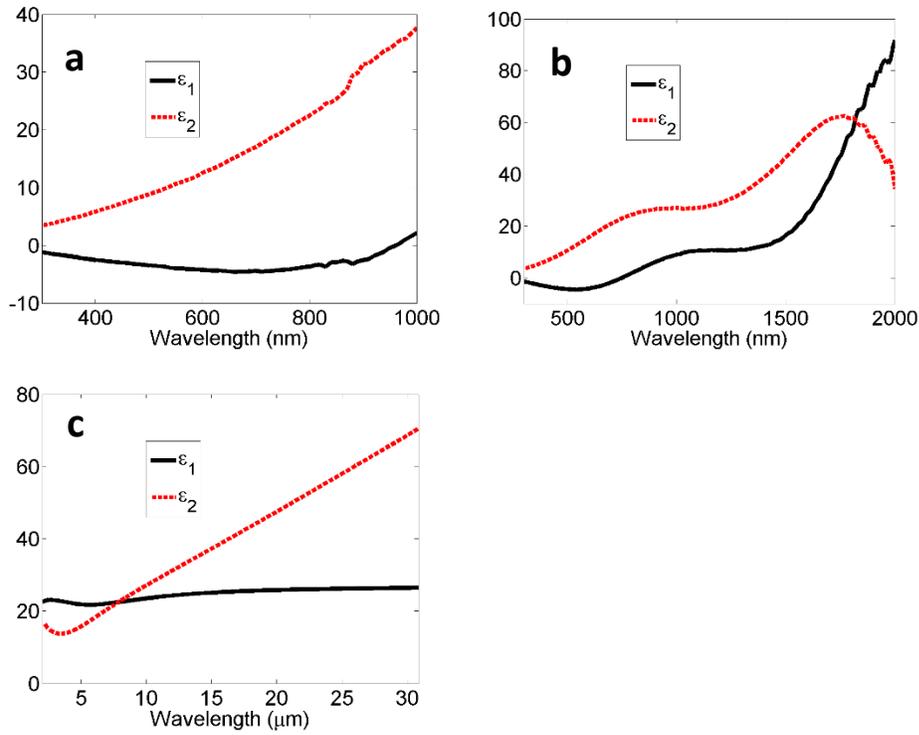

**Supplementary Figure 6 | TE dielectric functions. a,c**, Dielectric functions of bismuth telluride from 300-1000 nm (**a**) and 2-31 µm (**c**). **b**, Dielectric function of antimony telluride from 300-2000 nm.

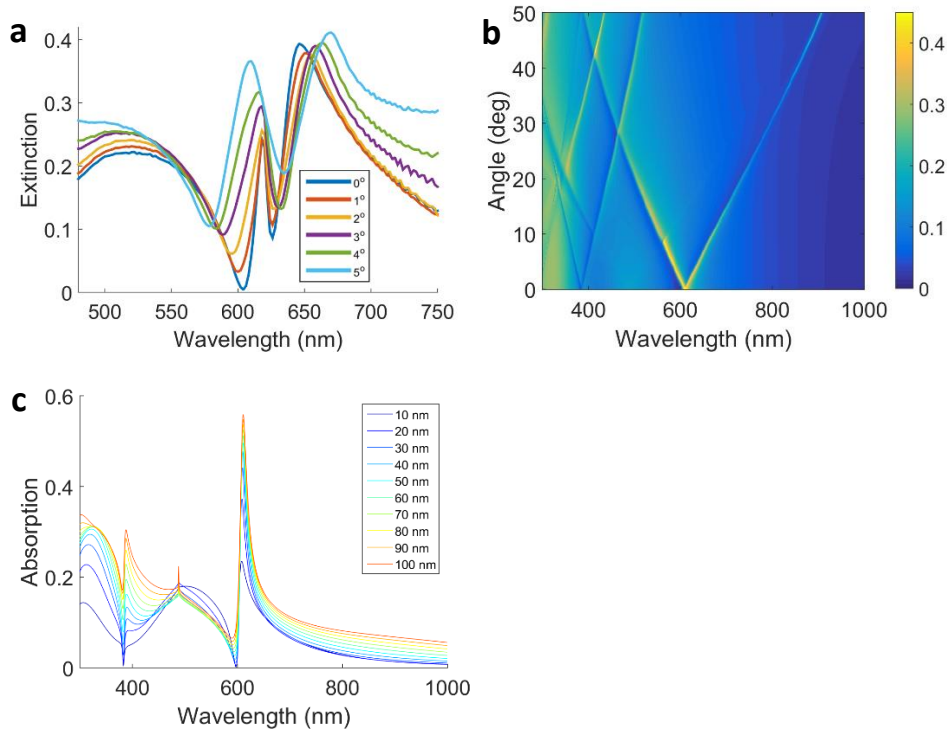

**Supplementary Figure 7 | Dependence of absorption spectra on incident illumination angle and wire height. a**, Measured extinction spectra for different angles of incidence. **b**, Full wave simulations of the angle dependence of 40 nm tall, 67 nm wide $Sb_2Te_3$ wires with a pitch of 488 nm on a 50 nm $SiO_2$ on 100 nm $SiN_x$ waveguide. We can see that even at 1 degree offset, the single peak splits into two. **c**, The dependence of wire height on absorption spectra, with pitch of 488 nm. We can see that absorption asymptotes to its maximum value around 40 nm and taller.

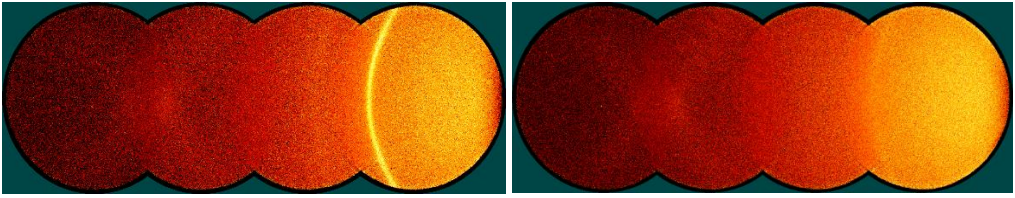

**Supplementary Figure 8 | XRD data.** XRD data of 100 nm of bismuth telluride (left) and 50 nm antimony telluride (right) shows very little crystallinity, as sputtered in experiments. Two-dimensional diffraction image frames were collected with frame centers set to 20, 40, 60 and 80 degrees in $2\theta$, from right to left, and then merged.

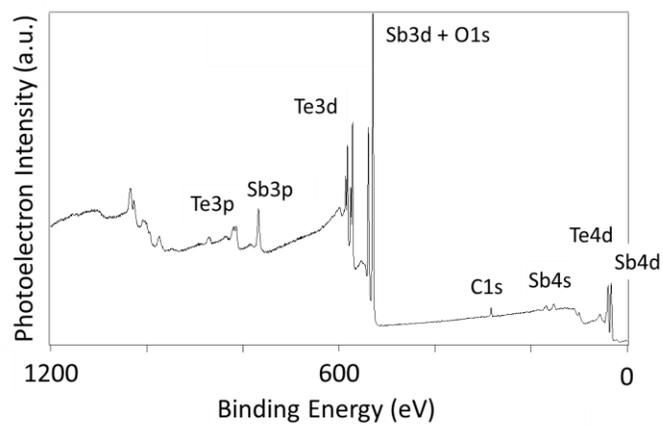

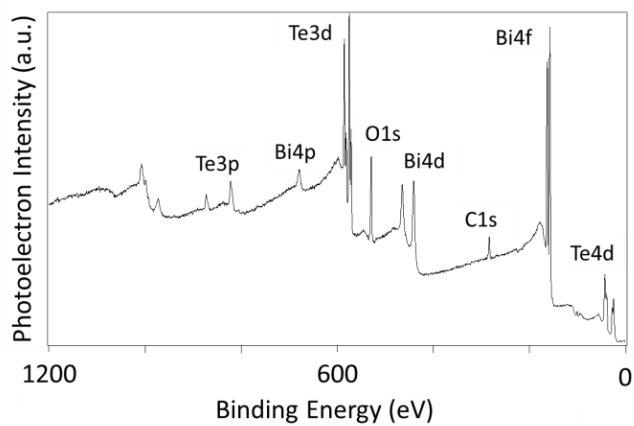

**Supplementary Figure 9 | XPS survey scans.** Sb2Te3 (top) and Bi2Te3 (bottom) samples.

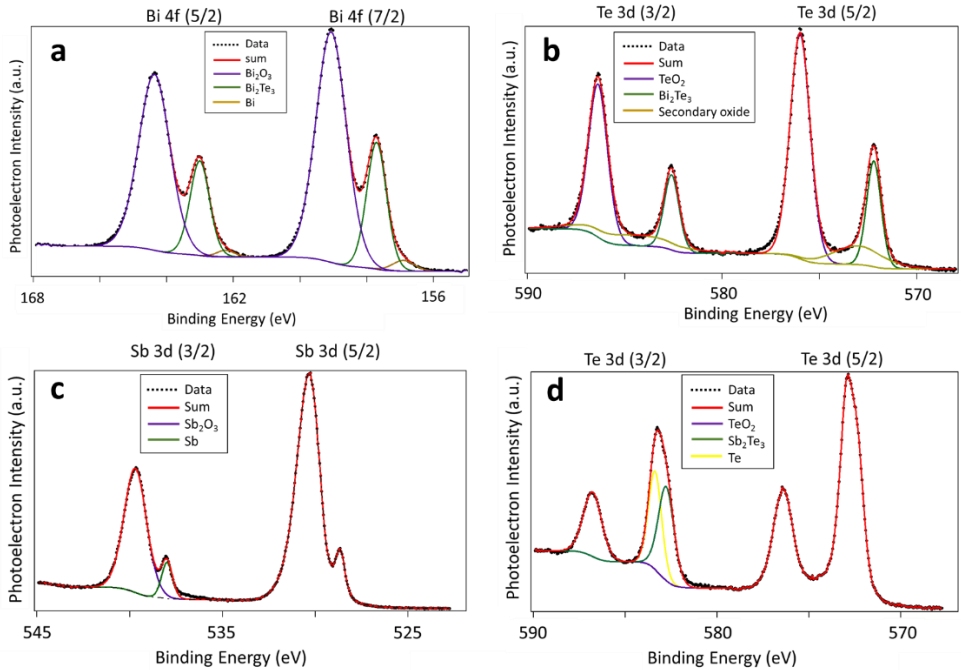

**Supplementary Figure 10 | Compositional analysis.** Detailed XPS data and fits for bismuth telluride peaks (**a-b**) and antimony telluride peaks (**c-d**) for our samples. **a**, Three components are visible in Bi 4f levels: the major components are $Bi_2O_3$, and $Bi_2Te_3$ (157.1 eV and spin-orbit pair at + 5.3 eV), with a small amount of elemental bismuth (156.6 eV). **b**, Te 3d level in bismuth-telluride is mostly $TeO_2$ and $Bi_2Te_3$ (582.3 eV, SO-splitting of 10.4 eV). **c**, Sb 3d levels show that most of the surface of antimony-telluride is in fact oxidized ($Sb_2O_3$), much more so in comparison with bismuth-telluride, with a measurable $Sb_2Te_3$ component (538 eV). **d**, Te 3d levels in antimony-telluride show the telluride, an oxide and elemental Te peaks.

| Figure | a (unitless) | $\omega_a$ (eV) | $W_a$ (eV) | $\omega_b$ (eV) | $W_b$ (eV) | q (unitless) | b (unitless) |
|---|---|---|---|---|---|---|---|
| S2/S3 | 0.13038 ± 0.000704 | 2.2174 ± 0.00014 | 0.020862 ± 0.000133 | 2.4448 ± 0.0067 | 0.48832 ± 0.00726 | -1.6986 ± 0.00943 | 0.82025 ± 0.0388 |
| S4/S5 – simulation | 0.58557 ± 0.00514 | 1.9725 ± 0.000991 | 0.041887 ± 0.000941 | 2.383 ± 0.0233 | 0.38067 ± 0.0203 | -1.2762 ± 0.0227 | 0.53022 ± 0.0578 |
| S4/S5 – experiment | 0.27396 ± 0.00968 | 1.9637 ± 0.00144 | 0.032595 ± 0.00151 | 2.3473 ± 0.0563 | 0.48954 ± 0.0991 | -2.185 ± 0.11 | 0.6564 ± 0.412 |

**Supplementary Table 1.** Fitting parameters for Supplementary equation (S7) for Supplementary Figs. 2-5. 95% confidence intervals are given.

| Absorption curve | Pitch$_s$ (nm) | Width$_s$ (nm) | $\Theta_s$ (deg) | Pitch$_f$ (nm) | Width$_f$ (nm) | $\Theta_f$ (deg) | Scaling |
|---|---|---|---|---|---|---|---|
| i | 567 | 97 | 0.5 | 560 | 97 | 0.5 | 0.71 |
| ii | 566 | 91 | 0.5 | 560 | 90 | 0.5 | 0.66 |
| iii | 511 | 119 | 0.5 | 507 | 102 | 0.6 | 1.06 |
| iv | 509 | 98 | 0.5 | 507 | 87 | 0.6 | 0.87 |
| v | 511 | 89 | 0.5 | 507 | 82 | 0.6 | 0.76 |
| vi | 452 | 131 | 1 | 455 | 102 | 1 | 0.99 |
| vii | 452 | 101 | 1 | 455 | 87 | 1 | 0.87 |

**Supplementary Table 2**. Comparison of experimental dimensions and illumination angle (Pitch$_s$, Width$_s$, $\Theta_s$) with best-fit simulation dimensions, illumination angle (Pitch$_f$, Width$_f$, $\Theta_f$) and scaling factor corresponding to Fig. 3d in the main text.

**Supplementary note 1. Thermal transport model calculations**

Making several approximations to equation (1) in the main text by volume and area averaging, we write:

$$\frac{1}{2}\omega\varepsilon''|\overline{E}|^2 V_{TE}$$
$$= eA_{air}\sigma(T^4 - T_0^4) - \kappa\nabla T A_{slice} + h(T - T_0)A_{air}$$
$$+ h_c(T - T_{sub})A_{interface}, \quad (S1)$$

where the optical absorption, radiation, conduction to the unilluminated portion of the TE, convection to air, and conduction to substrate are accounted for by each term, respectively. In Supplementary equation (S1), $\omega$ is the frequency of the incident light, $\varepsilon''$ is the imaginary part of the dielectric function of the TE material, $|\overline{E}|$ is the average magnitude of the electric field within the TE absorber material, $V_{TE}$ is the volume of the illuminated TE material, $e$ is the emissivity of the TE material, $A_{air}$ is the area of the TE exposed to the air, $\sigma$ is the Stefan-Boltzmann constant, $T_0$ is the initial temperature/cold end temperature/ambient air temperature, $\kappa$ is the thermal conductivity of the TE material, $A_{slice}$ is the cross-sectional area of the TE material separating the illuminated and unilluminated regions, $h$ is the heat transfer coefficient between the TE material and air, $h_c$ is the thermal boundary conductance between the TE material and the substrate, $T_{sub}$ is the temperature of the substrate near the TE material, and $A_{interface}$ is the area of the intersection of the TE material with the substrate.

The illuminated region for our structure is about 40 nm (height) by 100 nm (width) by 25 μm (length). We approximate the conduction term, $\kappa\nabla T A_{slice}$, by assuming to first order that $\nabla T \approx BT$, where $B$ is a constant, and is essentially the reciprocal of the total distance between the maximum temperature in the wire and the minimum temperature. From our 3D full wave simulation, we approximate that the largest temperature gradient occurs over about 1 micron, giving us $B \approx 1 \times 10^6$ m$^{-1}$. We use $\kappa =$ 1.2 W/mK, and $A_{slice}$ is the height multiplied by the width, or 4,000 nm$^2$, multiplied by two as heat is conducted away through both ends if the illuminated region.

The $A_{interface}$ is the width times the length, or 2.5 µm$^2$. $A_{air}$ is two times the height times the length plus the width times the length, or 4.5 µm$^2$. The other terms used are $T_0 = T_{sub} = 293$ K, $h = 100$ W/m$^2$K, $e = 0.4$, $h_c = 1 \times 10^5$ W/m$^2$K (this value is smaller to account for $T_{sub}$ being closer to the hot $T$ than to $T_0$), $\varepsilon'' = 0.4\varepsilon_0$, $\omega = 2.98 \times 10^{15}$ rad/Hz, and $|\overline{E}| = 8000$ V/m (from simulations corresponding to the power densities in our experiment). Solving Supplementary equation (S1) numerically, we find the steady state $\Delta T = 1.3$ K. From our experimental TEV, we expected a $\Delta T$ of 4 K on resonance, which is in the general validity range of our power flow model value.

**Supplementary note 2. Fano resonances**

Fano lineshapes are produced when a continuum of states interacts with discrete or narrow modes near the same energy, and appears in electronic circuits, nanophotonics, and atomic spectra alike[2]. In our specific case, we have a broad, Fabry-Perot (FP) resonance in our waveguide layers acting as the continuum background (radiative bright mode), and a narrow, waveguide mode (nonradiative dark mode) interacting with it. The effect is developed thoroughly in work by Gallinet et al.[1-3], which will be summarized here.

From the interaction between the bright (continuum) and dark (waveguide) modes, as Supplementary Fig. 1 outlines, we get a new resonance of the combined system, at a position equal to

$$\omega_a^2 = \omega_d^2 + \omega_d \Delta, \qquad (S2)$$

where $\omega_d$ is the resonant frequency of the dark mode and $\Delta$ is the shift away from this frequency due to coupling with the bright mode calculated explicitly in Gallinet[2]. The shape of the new resonance is asymmetric about the resonance, and is given by

$$\sigma_a(\omega) = \frac{\left(\frac{\omega^2 - \omega_a^2}{\Gamma} + q\right)^2 + b}{\left(\frac{\omega^2 - \omega_a^2}{\Gamma}\right)^2 + 1}, \qquad (S3)$$

where $q$ is an asymmetry term, $\Gamma$ is a width term equal to $2\omega_a W_a$ ($W_a$ is a width) for $\omega_a \gg W_a$, and $b$ is modulated damping term[2]. The asymmetry of this dark mode comes from a rapid phase asymmetry of ~pi across the resonance from the original dark mode, interfering with the symmetric phase difference in the bright mode across the resonance. On one side of the resonance, these bright and dark modes destructively interfere, and on the other side of the resonance they constructively interfere. The location of destructive interference on either side of the resonance depends on the sign of the phase difference between the dark and bright mode resonances, along with whether the loss is real or imaginary. This is expressed in the asymmetry term in Supplementary equation (S3), $q$, where

$$q = \pm \frac{(\omega_d^2 - \omega_b^2)}{2\Gamma_b \left(1 + \frac{\Gamma_i}{\Gamma_c}\right)}, \quad (S4)$$

where $\Gamma_b$ is equal to $2\omega_b W_b$ ($W_b$ is a width) for $\omega_b \gg W_b$, $\Gamma_i$ is intrinsic loss, and $\Gamma_c$ is coupling loss. $b$, the modulation damping parameter, is equal to

$$b = \frac{\left(\frac{\Gamma_i}{\Gamma_c}\right)^2}{\left(1 + \frac{\Gamma_i}{\Gamma_c}\right)^2} = \left(\frac{\Gamma_i}{\Gamma_c + \Gamma_i}\right)^2. \quad (S5)$$

The bright mode Fano formula, on the other hand, is pseudo-Lorentzian and therefore symmetric, given by

$$\sigma_b(\omega) = \frac{a^2}{\left(\frac{\omega^2 - \omega_b^2}{\Gamma_b}\right)^2 + 1}. \quad (S6)$$

From Gallinet[2], we find that the total optical response of the system comes from multiplying the Fano formulas of the bright mode with the dark mode modified by the bright mode, or

$$\sigma_{tot}(\omega) = \sigma_b(\omega)\sigma_a(\omega). \quad (S7)$$

Fitting our resonance to this formula, the frequency of the bright mode resonance can be extracted to determine its source, whether it be from plasma resonances of the wires, incident radiation, or FP resonances in the waveguide. In Supplementary Fig. 2 the Fano formula (Supplementary equation (S7)) is fit to our simulation data for a region near the resonance. All fitting parameters for Supplementary Figs. 2-5 are located in Supplementary Table 1.

The extracted bright mode from Supplementary Fig. 2 is shown in more detail in Supplementary Fig. 3, and is compared with the normalized magnitude of the electric field at the top of a 50 nm of $SiO_2$/100 nm of $SiN_x$ waveguide without wires due to FP resonances.

The minor misalignment of the bright mode and FP peak is likely caused by the grating itself altering the location of the bright mode, as the effective index of the photonic crystal made up of the grating plus the waveguide will be different than the index of the waveguide alone.

If we analyze our experimental spectra in this way and compare to simulation, we can compare differences quantitatively. Supplementary Figure 4 shows such a comparison. Supplementary Figure 5 plots the simulation and experimental bright modes from Supplementary Fig. 4 along with the FP resonance for a waveguide-only structure. If we compare the various values of the fitting parameters from Supplementary Table 1, we can note that $b$, the modulated damping term, is higher in the experiment than in the simulation. From Supplementary equation (S5), we can see that this indicates the intrinsic loss has a larger influence (or the coupling loss has a lesser influence) in the experiment than in the simulation. This could be attributed to $a$, the bright mode amplitude, having a higher magnitude in the simulation than in the experiment.

**Supplementary note 3. Dielectric functions of antimony telluride and bismuth telluride**

Supplementary Figure 6 shows the dielectric functions of our thin-film bismuth telluride and antimony telluride materials measured using J.A. Woollam Co. VASE and IR-VASE MARK II ellipsometers and WVASE software.

**Supplementary note 4. Angle/geometry dependence of guided mode resonance structure**

The absorption maxima of our wires structures is somewhat insensitive to wire height above 40 nm, and rather sensitive below, as illustrated via full wave simulations in Supplementary Fig. 7c. In contrast, the angle of incidence of light (away from normal incidence) has a very large effect on the absorption profile by splitting the single peak into two, as we can see in Supplementary Figs. 7a,b.

**Supplementary note 5. Absorption simulations and fitting**

The values for the width, pitch, and incident illumination angle for the experiment and simulations in Fig. 3d in the main text are shown in Supplementary Table 2. $Pitch_s$ and $Width_s$ of our fabricated structures was found via SEM imaging and were used in simulations for absorption spectra (red line in Fig. 3d). $\Theta_s$ was a best fit from simulation. $Pitch_f$, $Width_f$, and $\Theta_f$ were dimensions and incident illumination angle used in a simulation to best fit the experimental data. The best-fit simulation was multiplied by a scaling factor to further fit.

**Supplementary note 6. Bismuth telluride and antimony telluride compositional and structural analysis**

Two-dimensional XRD data were collected with a Bruker Discover D8 system, with a Vantec 500 detector and Cu Kα x-ray line produced by a microfocused IμS source, in a θ–2θ measurement. Supplementary Figure 8 shows results on thin films of bismuth-telluride and antimony-telluride, deposited using the same protocol as for the devices in the main text. We notice that in both cases the data show signatures of nanocrystallinity or perhaps even amorphous structure in the case of antimony-telluride.

X-ray photoemission spectroscopy (XPS) was performed on a Kratos Nova (Kratos Analytical Instruments), with monochromatized x-rays at 1486.6 eV and using a delay-line detector at a take-off angle of 35 degrees. The pressure during measurement was better than $5 \times 10^{-9}$ Torr, and the data were collected at 15 mA and 15 kV from an area of about 0.32 $mm^2$. Survey scans were collected at pass energy 160, and high-resolution scans at pass energy 10. Supplementary Figure 9 shows XPS survey spectra for our bismuth telluride and antimony telluride thin films, and due to surface sensitivity of the technique, these represent only the top few nanometers of the sample. Apart from expected surface oxidation and hydrocarbons from air, there are no other contaminants present that could affect the Seebeck coefficient. The stoichiometry of bismuth telluride and antimony telluride greatly affects the Seebeck coefficient[4-6], to the extreme that in the case of bismuth telluride, varying the amount of tellurium can change the carrier type from electron to holes over a small atomic percent composition change[6].

Using quantitative analysis based on Te 3d and Bi 4f levels, shown in Supplementary Fig. 10, we determined that the composition of our bismuth telluride was 42.5%:57.5%, Bi:Te for surface relative concentrations. This corresponds to a wt% of about 53.7% for the bismuth. Bottner et al.[4] found a Seebeck coefficient near -50 µV/K for cosputtered bismuth telluride on a non-heated substrate for our atomic concentrations. da Silva et al[5], working with thinner films than Bottner et al. but coevaporated on a heated substrate, found a similar result. Horne found that our compositions, in bulk, should be p-type. We also found that near stoichiometric compositions (such as ours) had a strong dependence on oxides, which is important in our systems as our XPS data shows large surface oxidation. However, our Hall measurements indicated the carriers in our bismuth telluride were electrons (and holes in antimony telluride), causing us to follow Bottner et al.'s findings.

The XPS measured composition of our 50 nm antimony telluride film was determined from Sb 3d 3/2 and Te 3d 3/2 peak areas (as identified to belong to the compound), and indicates a composition of 32%:68% Sb:Te. A large amount of antimony on the surface had oxidized. According to da Silva[5], this would correspond to a Seebeck coefficient of about 145 µV/K (p-type). Evidence further supports a p-type antimony telluride and an n-type bismuth telluride, as we saw the largest voltage produced when the wires were absorbing light directly on their junction (which would not occur for two p-type or two n-type thermoelectrics).

**Supplementary References**